\documentclass[conference]{IEEEtran}
\IEEEoverridecommandlockouts
\usepackage{cite}
\usepackage{amsmath,amssymb,amsfonts}
\usepackage{multirow}

\usepackage{graphicx}
\usepackage{textcomp}
\usepackage{xcolor}
\usepackage{comment}
\usepackage{booktabs}
\usepackage{graphicx}
\usepackage{tabularx}
\usepackage{pgfplots}
\usepackage{tablefootnote}
\usepackage{array}

\usepackage{wasysym}
\usepackage{tablefootnote}

\usepackage{hyperref}
\usepackage{pifont}
\usepackage{tikz}
\usetikzlibrary{shapes.geometric, arrows}
\newcommand{\cmark}{\ding{51}}%
\usepackage[caption=false]{subfig}
\usepackage{threeparttable}
\usepackage{lineno} 
\usepackage{multicol}
\usepackage{blindtext}
\usepackage{lscape}
\usepackage{tikz}
\usepackage{tikz-qtree}
\usetikzlibrary{patterns,arrows,positioning,calc,intersections,trees, chains, quotes, shapes.misc,
decorations.pathmorphing,positioning,decorations.pathreplacing,patterns,shapes.geometric, shapes.multipart,arrows.meta}
\def\BibTeX{{\rm B\kern-.05em{\sc i\kern-.025em b}\kern-.08em
    T\kern-.1667em\lower.7ex\hbox{E}\kern-.125emX}}
\begin{document}

\author{\IEEEauthorblockN{Adel Albshri\IEEEauthorrefmark{1}\IEEEauthorrefmark{2}, Bakri Awaji\IEEEauthorrefmark{3}\IEEEauthorrefmark{4} and Ellis Solaiman\IEEEauthorrefmark{5}}

\IEEEauthorblockA{\IEEEauthorrefmark{1}Newcastle University, School of Computing, UK, Email: a.albshri2@ncl.ac.uk}

\IEEEauthorblockA{\IEEEauthorrefmark{2}University of Jeddah, Saudi Arabia, 
Email: amalbeshri@uj.edu.sa}

\IEEEauthorblockA{\IEEEauthorrefmark{3}Newcastle University, School of Computing, UK, Email: b.h.m.awaji2@ncl.ac.uk}

\IEEEauthorblockA{\IEEEauthorrefmark{4}Najran University, Saudi Arabia, 
Email: balawaji@nu.edu.sa}

\IEEEauthorblockA{\IEEEauthorrefmark{5}Newcastle University, School of Computing, UK, Email: ellis.solaiman@newcastle.ac.uk}}

\title{Investigating the Requirements for Building a Blockchain Simulator for IoT Applications}
\maketitle
\thispagestyle{empty}
\pagestyle{empty} 
\begin{abstract}
The pervasiveness of the Internet of Things (IoT) has enabled the administration of a large number of intelligent devices. However, IoT is based on centralised models, which introduce a number of problems, such as a single point of failure and security risks. Blockchain may offer a viable option for addressing these concerns. Practically, both blockchain and IoT are complex technologies posing further challenges in assessing application performance. The availability of a reliable simulation environment for Blockchain based IoT applications would be a major aid in the development and evaluation of such applications. Our research has found that  currently there are no simulators with a comprehensive set of features, for the development and evaluation of blockchain based IoT applications, which is the main motivation for our work. The purpose of this study is to gather the opinions of experts regarding the creation of a simulation environment for IoT based blockchain applications. To do this, we utilise two separate investigations. First, a questionnaire is developed to ensure that the development of such simulation software would be of significant use. Second, interviews with participants are performed to gain their perspectives on the primary issues they face with blockchain-based IoT applications. In addition, the interviews focused on collecting the perspectives of participants on how blockchain may improve IoT and how to identify blockchain's applicability in IoT. Our findings demonstrate that the participants had a great deal of confidence in blockchain to resolve IoT issues. However, they lack the tools necessary to assess this concept. This highlights their requirement for a simulator to analyse the integration of blockchain and IoT.

\end{abstract}


\section{Introduction}
The Internet of Things (IoT) is a pervasive network of interconnected things (For example, sensors, actuators, smart TV and smart cars) that can connect and exchange data with other devices or users. Formally, IoT consists of a number of essential components~\cite{Zaidan2018}. First, things sense an environment’s parameters, and  transmit their values to local edge devices or to central Cloud servers for processing. This occurs via the second component which is the gateways. These gateways are the mediators among things, edge devices, and the cloud, which support the required connectivity, security, as well as data flow~\cite{ejaz2020performance}. The third and fourth components are local edge devices for small scale data processing, and/or cloud infrastructure, which comprises huge pools of virtualised resources (For example, storage and processing devices) possessing ultra-fast processing and analytical power~\cite{zhong2015study}. 

Certain challenges have emerged with IoT security, especially in privacy which hinders the adoption of IoT applications. Specifically, central IoT servers are responsible for managing users' sensitive data, which, in turn, could affect privacy and cause privacy violations~\cite{vangala2020smart}. Internet of Things applications are wide ranging, including health based applications, environmental and Geospatial applications. Many of these applications involve the exchange of sensitive and private data. An important question therefore poses itself; how can we ensure that this data has not been altered tampered with, or misused? This question is particularly pertinent to the centralised IoT architectures, which have to deal with the single point of failure problem. Also a centralised architecture for IoT applications where potentially huge volumes of data are being exchanged may have a negative impact on the performance of such applications, slowing them down to levels that could make them dangerous, for example in cases where a hospital does not get patient data at critical moments~\cite{li2021blockchain}. Thus, there is a need to explore decentralised models for implementing IoT applications. The peer-to-peer (P2P) model for handling large exchanges between IoT devices has the potential to fundamentally reduce the cost of employing servers~\cite{hosseinian2019smart}. It also distributes processing tasks over a larger number of devices. Consequently, failure of any single device in the network will not result in failure, thus meeting fault tolerance requirements. Additionally, the immense cost of servers, their operating systems and maintenance is averted by way of such P2P networks \cite{delgado2018cryptocurrency}. Nevertheless, adopting this P2P model will come with its well-known security issues~\cite{washbourne2015survey}. It is here where blockchain, known to be an extended secure P2P network~\cite{li2018block} can be very effective. 

Blockchain is widely known to be a tamper-proof and secure technology. Bitcoin, the digital currency of P2P, exhibits the most popular example of employing blockchain technology~\cite{nakamoto2008bitcoin}. Typically, blockchain is capable of securing data by means of transparency where all network participants share the same data that can only be updated through consensus. Moreover, blockchain allows us to trace data back to its initiator in an efficient and low-cost strategy. Therefore there has been increased interest in exploring the integration of blockchain technology into IoT Networks. Doing so would potentially culminate in the development of fully distributed digital networks that are trustworthy and more efficient~\cite{memon2019dualfog}. However, both IoT and Blockchain technologies are highly complex, and so are many of their potential applications. Therefore the development of accurate and effective simulation tools that can model and evaluate such applications before they are implemented in the real world, would be of huge benefit. 

Simulators investigate a system's parameters and behaviour \cite{Haverkort1998}. They are especially useful with complex systems that need to be examined before implementation~\cite{ferretti2020ethereum}\cite{harbers2018conceptual}. Blockchain and IoT are good examples that can benefit from simulation because blockchain and IoT consist of various interconnected layers \cite{zhu2019applications}\cite{soumyalatha2016study}.  Simulators can substantially reduce the financial costs needed to deploy real blockchain or IoT systems. Also, simulators allow investigating a system's performance under different configuration setups.


In this context, the fundamental objective of this research is to develop a simulator for analysing the performance of integrated IoT blockchain systems. Motivated by this, the purpose of the current study is to gather the views and viewpoints of specialists in the area about the design of an adequate simulator. This was accomplished via the use of a questionnaire and interviews. The findings indicate a high level of support for the design of an Internet of Things blockchain simulator.

The remainder of this paper is organised as follows. Section~\ref{Relatedwork} covers the related work. Section~\ref{objective} presents the objectives of the study. Section~\ref{methods} presents the paper’s adopted methods. In Sections~\ref{results} and~\ref{discussion}, the results are presented and discussed. Section~\ref{recommendation} presents the study’s proposals. The paper finishes with the conclusions in section~\ref{conclusion}.

\section{Related work}
\label{Relatedwork}

Over the last few years, Blockchain and IoT have been the focus of much research. In the literature, several attempts have been made to simulate blockchain~\cite{paulavivcius2021systematic} or IoT applications~\cite{markus2020survey} separately. VIBES (Visualisations of Interactive Blockchain Extended Simulations) \cite{Stoykov2017} was proposed as a configurable blockchain simulator to enable end-users to perceive empirical insights and analytics about blockchain networks. VIBES can simulate blockchain systems and mimics the effect of specific parameter changes on the system. The merits of VIBES are twofold. First, VIBES is a scalable simulator as it can simulate systems with thousands of interacting nodes. Second, VIBES is a fast simulator able to provide fast simulation results. Faria and Correia~\cite{Faria2019} proposed a discrete-event blockchain simulator referred to as BlockSim that is a framework assisting in designing, implementing and evaluating blockchains. It can evaluate the implementation of different blockchains that are rapidly modelled and simulated. Therefore, BlockSim is characterised as a dynamic simulator able to simulate systems over a certain time interval. Yet another attempt referred to as BlockSim is proposed by Alharby and van Moorsel~\cite{Alharby2019} that implements proof of work (PoW) as a consensus algorithm for making the agreements about the blockchain's state. Moreover, as a discrete-event simulator, BlockSim can test the effect of different parameter configurations on the system’s performance. Another simulator BlockSIM \cite{Pandey2019} is a resilient open-source blockchain simulator that enables blockchain designers to evaluate the performance of their designed private blockchains. The contribution of BlockSIM is twofold. First, it accurately models the stability of the system. Second, it accurately simulates the transaction throughput concerning a given scenario. It can optimise the system's parameters which, in turn, allows for testing various scenarios needed for the build-up of its chains. 
 
Several attempts have been made to simulate IoT ecosystems and their applications. A cloud layer is normally significant for a wide range of IoT applications; therefore, cloud simulators are widely used for simulating IoT applications. The most popular and widely used is the CloudSim toolkit \cite{calheiros2011cloudsim} in which tasks are created in the form of cloudlets to be processed using virtual machines in the cloud environment. Moreover, it is mainly designed to simulate the discrete-event scenarios while implementing a five-layer structure. An interesting aspect of CloudSim is its ability to model CPU power consumption to shed light on bandwidth and delay parameters. Due to its success, an improved extended version is introduced and referred to as CloudAnalyst~\cite{CloudAnalyst}. CloudAnalyst extends the core of CloudSim while adding a set of features to investigate the effect of different configurations on the system's performance. A prominent simulator for modelling applications on the Edge of IoT networks is iFogSim~\cite{gupta2017ifogsim} which is an extension of the CloudSim simulator. As an edge layer-dependent simulator, it can simulate real systems that consider the different aspects ranging from sensing to processing the data. The main contribution of this simulator is the simulation of the physical layer. In particular, it can model the physical component of systems.

To our knowledge, none of the simulators mentioned above focuses on simulating IoT scenarios (IoT applications that run on multiple IoT layers, including sensors, edge devices, communication networks, and cloud). This motivated the development of IoTSim~\cite{zeng2017iotsim}. IoTSim is built over the core of CloudSim to support the task of IoT and big data simulation. IoTSim follows a three-layer architecture (perception, network and application layer). These layers are integrated with the three layers of CloudSim (storage, big data processing and user code). An important point in this simulator is using the MapReduce approach, one of the big data handling approaches. From the practical viewpoint, this is done through two separated functions: MapCloudlet and ReduceCloudLet. Finally, IoTSim-Osmosis~\cite{ALWASEL2021101956} is a framework that supports testing and validating IoT applications using the principle of osmotic computing. It is mainly designed to simulate complex IoT applications while being deployed on heterogeneous edge, cloud and SDN environments. Despite the many great efforts done so far, none of the previously mentioned simulators, summarised in Table \ref{summarize}, focus on simulating the integration of blockchain with IoT.

\begin{table*}
\centering
\caption{A summary of the related work simulators along with their main features.}
\label{summarize}
\scalebox{0.8}{
\begin{tabular}{lccccccccc||ccc} 

\hline
 Simulator&Simulator &Programming&Core&Simulator
& \multicolumn{8}{c}{\textbf{Features}} \\\cline{6-13}
&Scope& language&&Type & \multicolumn{5}{c}{End to end IoT layers} & \multicolumn{3}{c}{Blockchain layers}\\\cline{6-13}
&&&&&IoT&Edge&Network&Network&Cloud&Network &Consensus &Data \\ 
&&&&&device&&communication&Protocol&&&&\\ \hline
VIBES~\cite{Stoykov2017}&Blockchain&Scala&N/A&Discrete-event&&&&&&&\cmark&\cmark\\ 
BlockSim~\cite{Faria2019}&Blockchain&Python&N/A&Discrete-event&&&&&&\cmark&\cmark&\cmark\\ 
BlockSim~\cite{Alharby2019}&Blockchain&Python&N/A&Discrete-event&&&&&&&\cmark&\cmark\\

BlockSIM~\cite{Pandey2019}&Blockchain&Python&N/A&Discrete-event&&&&&&&\cmark&\cmark \\ \hline \hline
CloudSim~\cite{calheiros2011cloudsim}&Cloud&Java&N/A&Discrete-event&&&&&\cmark&&&\\
CloudAnalyst~\cite{CloudAnalyst}&Cloud&Java&CloudSim&Discrete-event&&&&\cmark&\cmark&&&\\
iFogSim~\cite{gupta2017ifogsim}&Edge&Java&CloudSim&Discrete-event&\cmark&\cmark&&&&&&\\
IoTSim~\cite{zeng2017iotsim}&IoT&Java&CloudSim&Discrete-event&\cmark&&&\cmark&&&&\\

IoTSim-Osmosis~\cite{ALWASEL2021101956}&End to End IoT&Java&CloudSim&Discrete-event&\cmark&\cmark&\cmark&\cmark&\cmark&&&\\  \hline 

\end{tabular}

}
\end{table*}

\section{Objectives}
\label{objective}
This paper aims to obtain the opinions and perspectives of research participants regarding blockchain's potential contributions to IoT. For example, enabling IoT data transparency and security. Once the participants' thoughts are gathered, the proposed system’s requirements can be established as well as the required tools and mechanisms. This process is described in relation to a number of objectives, as follows:
\begin{enumerate}
    \item To gather the required information from experts in the field regarding:
    \begin{enumerate}
        \item The usage of IoT in our daily life.
        \item The most commonly used blockchain types.
        \item The IoT data that should be stored on blockchain.
        \item The consensus algorithms required for the simulator.
        \item The users' needs as regards the blockchain log.
        \item The possibility of using IoT nodes as blockchain nodes.
    \end{enumerate}
    \item To provide analytical information regarding:
    \begin{enumerate}
        \item Participants' opinions about having an integrated blockchain IoT simulator.
        \item Participants' opinions on modelling various types of blockchain in the simulator.
    \end{enumerate}
    \item To design a simulator to validate the integrated blockchain IoT systems.
\end{enumerate}

\section{Method}
\label{methods}
\subsection{Participants}
This paper employed a sequential explanatory design methodology~\cite{creswell2017designing} comprising a questionnaire and interviews. Overall, there were 25 participants who represented the target sample of individuals with knowledge of computer science, with a specific specialisation in IoT and/or blockchain.


\subsection{Research tools}
An online questionnaire with nine closed-ended questions was prepared using the SurveyMonkey website. The questionnaire was distributed to the participants. This was followed up by online interviews using Zoom app, with a set of participants who provided their consent to participate. By the end of the interview, participants had the opportunity to complete a form with a set of open questions. This enables qualitative data to be collected to provide a high level of analysis. For statistical analysis, the Cronbach’s Alpha~\cite{nunnaly1978psychoneric} is calculated using SPSS for the 9 questions resulting in 0.796 data consistency. Evidently, the value exceeds 0.5, which, in turn, indicates the high reliability and consistency of gathered information.

\subsection{Research procedures}
First, it was necessary to gather quantitative numerical data through a questionnaire~\cite{Pope2000}, with the aim of developing robust conclusions. Second, qualitative data was gathered through interviews with various participants, using a set of open questions~\cite{Gill2008}. The first approach, the questionnaire, was disseminated to approximately 25 participants, all of whom were IoT and blockchain, developers/researchers. With 25 active participants, the statistical analysis was undertaken using SPSS to understand the participants' attitudes regarding blockchain features. To more effectively communicate the idea, the data analysis results as numeric values are presented in descriptive graphical format. The question responses were provided on a Likert scale from 1 (`strongly disagree') to 5 (`strongly agree'). The findings, presented in the figures, are displayed in the~\ref{QuestionnaireResult}. Regarding the second data collection approach of the interviews, these were undertaken online with six participants who responded to a set of open questions. An in-depth description of this process is presented in~\ref{InterviewsResult} section. 

\section{Findings}
\label{results}
\subsection{Questionnaire}
\label{QuestionnaireResult}
\begin{figure}[htbp]
\centering
\begin{tikzpicture}  
\begin{axis}  
[  
    ybar,  
    title = {\footnotesize Q1: To what extend are you familiar with IoT?},
    ylabel={ \footnotesize Frequency}, 
    xlabel={\footnotesize  Familiarity level},  
    symbolic x coords={Low, Moderately low, Moderately, Moderately high, High}, enlargelimits=true,x tick label style={font=\footnotesize,text width=1cm,align=center},
    y label style={below=0.5mm},
    x label style={below=2mm},
    bar width=0.8cm,
    xtick=data,  
     nodes near coords, 
    nodes near coords align={vertical},  
    bar shift=0,
          height=5.7cm,
          width=8.5cm,
          style={xshift=30pt,yshift=0pt}
    ]  
\addplot +[black, pattern=north east lines] coordinates {(Low,2) (Moderately low,6) (Moderately,8) (Moderately high,5) (High,4) };  
  
\end{axis}  
\end{tikzpicture}  
    \caption{Participants' familiarity with the IoT.}
	\label{IoTfamiliarity}
\end{figure}
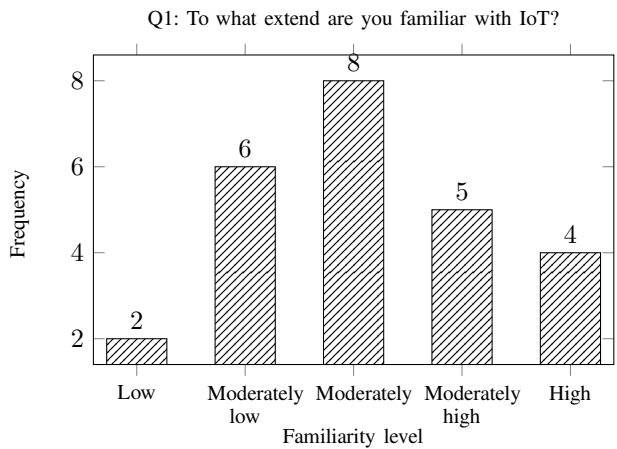
The questionnaire begins by asking questions to determine the participants’ familiarity with the IoT. Specifically, the participants were asked, \textit{``To what extent are you familiar with IoT?"} In this case, 25 answers presented in Figure~\ref{IoTfamiliarity} were received from participants regarding their familiarity. The figure shows that the majority (eight participants, 32\%) are moderately aware of the IoT, while six participants (24\%) have moderately low familiarity with the IoT. Moreover, four participants (16\%) are highly aware of the IoT, with another five participants (20\%) possessing moderately high awareness of the IoT. Conversely, the least number (two participants, 8\%) are completely unaware of the IoT. Accordingly, the selected participants are a good fit because the majority (moderate and higher) are aware of the IoT.

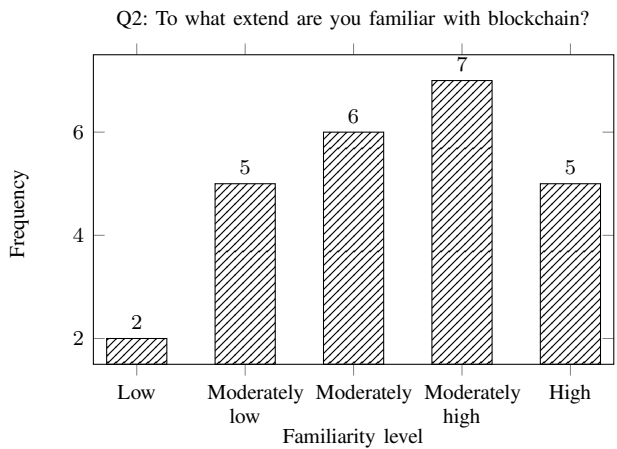
\begin{figure}[htbp]
    \centering
\begin{tikzpicture}  
  
\begin{axis}  
[  
    ybar,  
    title = {\footnotesize Q2: To what extend are you familiar with blockchain?},
    ylabel={\footnotesize Frequency}, 
    xlabel={\footnotesize Familiarity level},  
    symbolic x coords={Low, Moderately low, Moderately, Moderately high, High}, 
    enlargelimits=true,x tick label style={font=\footnotesize,text width=1cm,align=center},
    y label style={below=0.5mm},
    x label style={below=2mm},
    bar width=0.8cm,
    xtick=data,  
     nodes near coords, 
    nodes near coords align={vertical},  
    bar shift=0,
          height=5.7cm,
          width=8.5cm,
          style={xshift=0pt,yshift=0pt,anchor=north,font=\footnotesize}
    ]  
\addplot +[black, pattern=north east lines]coordinates {(Low,2) (Moderately low,5) (Moderately,6) (Moderately high,7) (High,5) };  
\end{axis}  
\end{tikzpicture}  
    \caption{Participants' familiarity with Blockchain.}
	\label{Blockfamiliarity}
\end{figure}

As we are discussing two technologies, we also examined their familiarity with blockchain to feel more confident about the participants’ answers. Thus, the participants were asked, \textit{``To what extent are you familiar with blockchain?"} In this instant, 25 responses, presented in Figure~\ref{Blockfamiliarity}, were received from the respective participants regarding their familiarity. The figure suggests that the majority (seven participants, 28\%) possess moderately high awareness of blockchain, while six participants (24\%) are very familiar with blockchain. The least number of participants (two, 8\%) are completely unaware of blockchain, while five participants (20\%) possess moderately low awareness of blockchain. Additionally, six participants (24\%) are moderately aware of blockchain. 
\begin{figure}[htbp]
    \centering
\begin{tikzpicture}  
  
\begin{axis}  
[  
    ybar, title style = {text width = 8cm,align = center},
    title = {\footnotesize Q3: From my perspective, there may be an expansion of blockchain technology along with IoT in the future.},
    ylabel= {\footnotesize Frequency}, 
    xlabel={\footnotesize Agreeing level},  
    symbolic x coords={Strongly disagree, Disagree, Neutral, Agree, Strongly agree}, 
    bar width=1cm,
    xtick=data,  
     nodes near coords, 
    nodes near coords align={vertical},  
enlargelimits=true,x tick label style={font=\footnotesize,text width=1cm,align=center},
    y label style={below=0.5mm},
    x label style={below=2mm},
    bar width=0.8cm,          
          height=5.7cm,
          width=8.5cm,
          style={xshift=0pt,yshift=0pt,anchor=north,font=\footnotesize, fill=red}
    ]  
\addplot +[black, pattern=north east lines] coordinates {(Strongly disagree,1) (Disagree,7) (Neutral,3) (Agree,6) (Strongly agree,8) };  
  
\end{axis}  
\end{tikzpicture}  
    \caption{Participants' thoughts about the IoT's integration with blockchain.}
	\label{thoughts}
\end{figure}
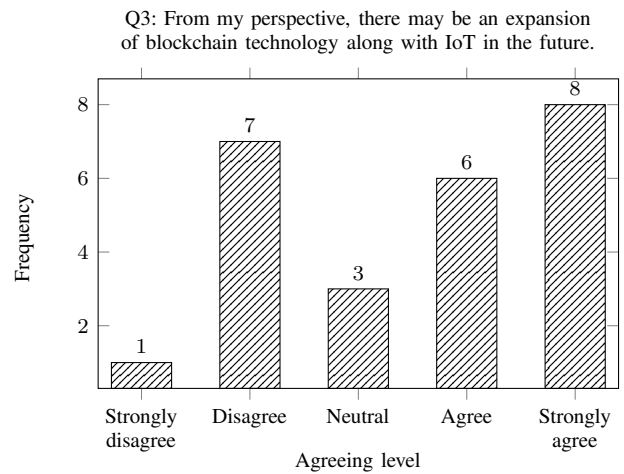
Similar to the participants’ familiarity with the IoT, the selected participants are a good fit, given that the majority are aware of blockchain. Subsequently, the participants were asked, \textit{``if they believe that there will be an expansion of blockchain with IoT in the future"} all 25 participants responded to this question, with their responses presented in Figure~\ref{thoughts}. It was established that the majority (eight participants, 32\%) highly agreed with this point. Moreover, six participants (24\%) expressed moderately high agreement with the idea. In total, 11 participants (44\%) either moderately or completely disagreed. 

\begin{figure}[htbp]
    \centering
\begin{tikzpicture}  
  
\begin{axis}  
[  
    ybar,title style = {text width = 10cm,align = center},
    title = {\footnotesize Q4: Do you think there is a need for an IoT blockchain simulator that helps developers to systematically adjust the system's configurations?},
    ylabel={\footnotesize Frequency}, 
    xlabel={\footnotesize Agreeing level},  
    symbolic x coords={Strongly disagree, Disagree, Neutral, Agree, Strongly agree}, 
    bar width=1cm,
    xtick=data,  
     nodes near coords, 
    nodes near coords align={vertical},  
enlargelimits=true,x tick label style={font=\footnotesize,text width=1cm,align=center},
    y label style={below=0.5mm},
    x label style={below=2mm},
    bar width=0.8cm,         
          height=5.7cm,
          width=8.5cm,
          style={xshift=0pt,yshift=0pt,anchor=north,font=\footnotesize}
    ]  
\addplot +[black, pattern=north east lines] coordinates {(Strongly disagree,2) (Disagree,4) (Neutral,2) (Agree,8) (Strongly agree,9) };  
  
\end{axis}  
\end{tikzpicture}  
    \caption{Participants' thoughts about having an integrated IoT blockchain simulator.}
	\label{configuation}
\end{figure}
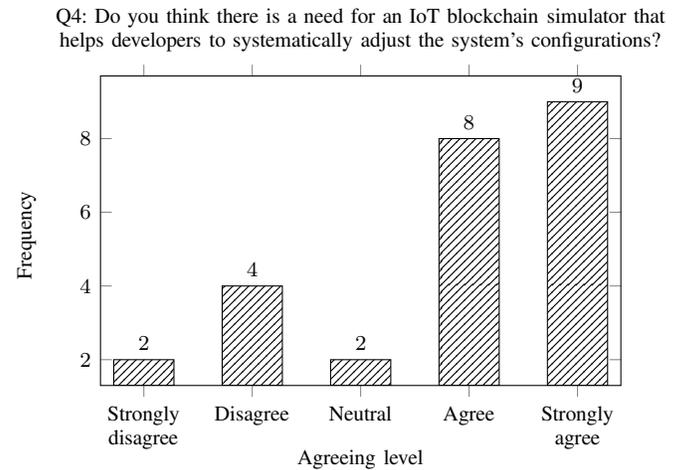

Following this, participants were asked, \textit{``What are your thoughts regarding the need to have an IoT blockchain simulator for helping developers with adjusting the system's configurations?”} all 25 participants provided their responses, summarised in Figure~\ref{configuation}. As is apparent from this figure, nine participants (36\%) strongly agreed with this notion, while eight participants (32\%) agreed with this concept. In total 10 participants (32\%) were either neutral or completely disagreed.
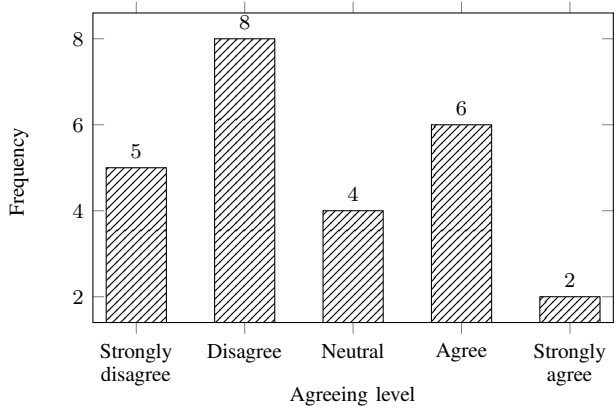
\begin{figure}[htbp]
    \centering
\begin{tikzpicture}  
  
\begin{axis}  
[  
    ybar,  
    title = {\footnotesize Q5: Do you agree that all IoT data should be stored in the blockchain?},
    ylabel={\footnotesize Frequency}, 
    xlabel={\footnotesize Agreeing level},  
    symbolic x coords={Strongly disagree, Disagree, Neutral, Agree, Strongly agree}, 
    bar width=1cm,
    xtick=data,  
     nodes near coords, 
    nodes near coords align={vertical},  
enlargelimits=true,x tick label style={font=\footnotesize,text width=1cm,align=center},
    y label style={below=0.5mm},
    x label style={below=2mm},
    bar width=0.8cm,
          height=5.7cm,
          width=8.5cm,
          style={xshift=0pt,yshift=0pt,anchor=north,font=\footnotesize}
    ]  
\addplot +[black, pattern=north east lines] coordinates {(Strongly disagree,5) (Disagree,8) (Neutral,4) (Agree,6) (Strongly agree,2) };  
  
\end{axis}  
\end{tikzpicture}  
    \caption{Participants' thoughts about storing all of the IoT data in the blockchain.}
	\label{IoTdata}
\end{figure}

Given that the participants are domain experts, we seized the opportunity to obtain their perspectives regarding storing the IoT data in blockchain. Therefore, the participants were asked, \textit{``Do you agree that all IoT data should be stored in the blockchain?”} the participants’ responses are presented in Figure~\ref{IoTdata}. Evidently, the majority disagreed with this statement (13 participants either disagreed or strongly disagreed). The reason behind this may be the different scenarios of use concerning IoT with blockchain. Alternatively, the least number of participants expressed agreement with this statement (eight participants either agreed or strongly agreed), while two participants felt neutral regarding this statement. 

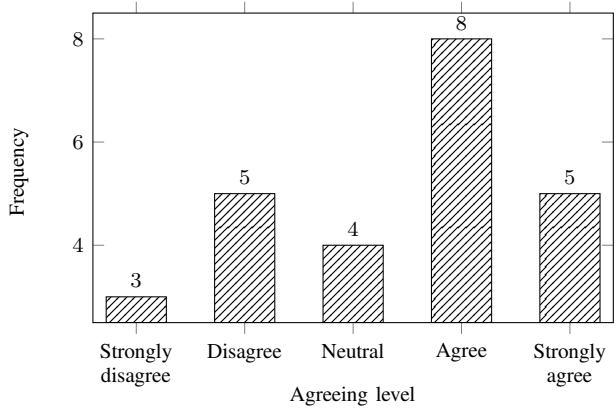
\begin{figure}[htbp]
    \centering
\begin{tikzpicture}  
  
\begin{axis}  
[  
    ybar,title style = {text width = 9cm,align = center},
    title = {\footnotesize Q6: Do you agree having an IoT blockchain simulator that makes use of different types of consensus algorithm?},
    ylabel={\footnotesize Frequency}, 
    xlabel={\footnotesize Agreeing level},  
    symbolic x coords={Strongly disagree, Disagree, Neutral, Agree, Strongly agree}, 
    bar width=1cm,
    xtick=data,  
     nodes near coords, 
    nodes near coords align={vertical},  
enlargelimits=true,x tick label style={font=\footnotesize,text width=1cm,align=center},
    y label style={below=0.5mm},
    x label style={below=2mm},
    bar width=0.8cm,                 
            height=5.7cm,
          width=8.5cm,
          style={xshift=0pt,yshift=0pt,anchor=north,font=\footnotesize}
    ]  
\addplot +[black, pattern=north east lines] coordinates {(Strongly disagree,3) (Disagree,5) (Neutral,4) (Agree,8) (Strongly agree,5) };  
  
\end{axis}  
\end{tikzpicture}  
    \caption{Participants' thoughts about having multiple consensus algorithm in the simulator.}
	\label{consensus}
\end{figure}
Consensus algorithms are of considerable importance in blockchain because they are used to reach a common agreement (consensus) on the current state of the ledger data. They also enable unknown peers to be trusted in a distributed computing environment. Thus, there is a need to establish participants’ needs relating to this. A. Accordingly, the participants were asked, \textit{``What are your thoughts on having multiple consensus algorithms in the simulator?"} the participants’ responses to this question are summarised in Figure~\ref{consensus}. Considering the data more closely, it is apparent that the majority (eight participants, 32\%) agreed with this notion. Furthermore, five participants (20\%) strongly agreed with the idea. In total, 11 participants (38\%) expressed either moderate or complete disagreement. 

\begin{figure}[htbp]
    \centering
\begin{tikzpicture}  
  
\begin{axis}  
[  
    ybar,  title style = {text width = 9cm,align = center},
    title = {\footnotesize Q7: Do you prefer having the flexibility to investigate the detailed log information for every transactions?},
    ylabel={\footnotesize Frequency}, 
    xlabel={\footnotesize Agreeing level},  
    symbolic x coords={Strongly disagree, Disagree, Neutral, Agree, Strongly agree}, 
    bar width=1cm,
    xtick=data,  
     nodes near coords, 
    nodes near coords align={vertical},  
enlargelimits=true,x tick label style={font=\footnotesize,text width=1cm,align=center},
    y label style={below=0.5mm},
    x label style={below=2mm},
    bar width=0.8cm,               
    height=5.7cm,
          width=8.5cm,
          style={xshift=0pt,yshift=0pt,anchor=north,font=\footnotesize}
    ]  
\addplot +[black, pattern=north east lines] coordinates {(Strongly disagree,5) (Disagree,3) (Neutral,5) (Agree,8) (Strongly agree,4) };  
  
\end{axis}  
\end{tikzpicture}  
    \caption{Participants' thoughts about the ability to investigate the log.}
	\label{log}
\end{figure}
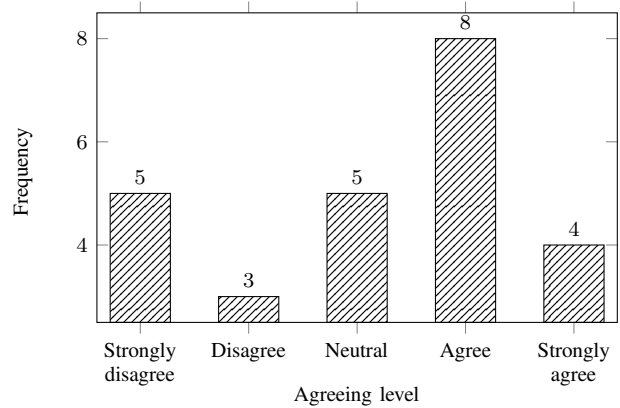

Considering blockchain in greater depth, it is essential to determine the participants’ perspectives regarding investigating the log. This is crucial because it provides the opportunity to compute system latency and throughput. Accordingly, the participants were asked for their opinions concerning investigating the log file. The participants’ responses to this question are presented in Figure~\ref{log}. The significant point is that the majority (12 participants) either strongly agreed or agreed with this idea. Additionally, five participants expressed neutrality concerning the statement. Meanwhile, eight participants in total expressed either moderate or complete disagreement. 

\begin{figure}[htbp]
    \centering
\begin{tikzpicture}  
  
\begin{axis}  
[  
    ybar,title style = {text width = 9cm,align = center},
    title = {\footnotesize Q8: Do you agree using IoT edge devices like Raspberry pi as the blockchain nodes?},
    ylabel={\footnotesize Frequency}, 
    xlabel={\footnotesize Agreeing level},  
    symbolic x coords={Strongly disagree, Disagree, Neutral, Agree, Strongly agree}, 
    bar width=1cm,
    xtick=data,  
     nodes near coords, 
    nodes near coords align={vertical},  
enlargelimits=true,x tick label style={font=\footnotesize,text width=1cm,align=center},
    y label style={below=0.5mm},
    x label style={below=2mm},
    bar width=0.8cm,         
    height=5.7cm,
          width=8.5cm,
          style={xshift=0pt,yshift=0pt,anchor=north,font=\footnotesize}
    ]  
\addplot +[black, pattern=north east lines] coordinates {(Strongly disagree,3) (Disagree,6) (Neutral,4) (Agree,5) (Strongly agree,7) };  
  
\end{axis}  
\end{tikzpicture}  
    \caption{Participants' thoughts about using IoT edge devices as blockchain nodes.}
	\label{pi}
\end{figure}
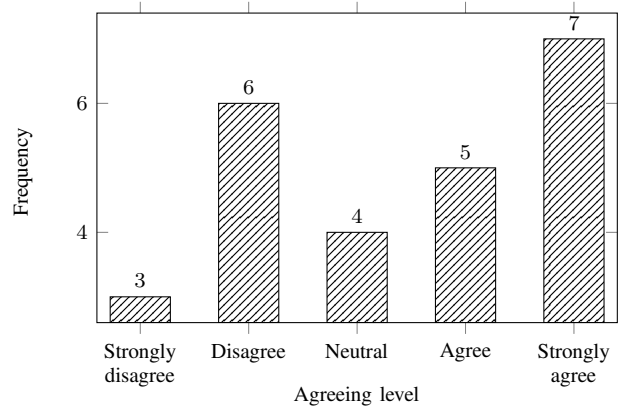

Subsequently, the participants were asked about using IoT devices as blockchain nodes. The participants’ responses to this question are presented in Figure~\ref{pi}, which presents their overall positive perspectives regarding this statement. Ultimately, most participants either strongly agreed (seven participants, 28\%) or agreed (five, 20\%) with the statement. In contrast, a total of nine participants (36\%) either strongly disagreed or disagreed with this notion. Lastly, six participants (24\%) expressed neutrality regarding this notion. 

\begin{figure}[htbp]
    \centering
\begin{tikzpicture}  
  
\begin{axis}  
[  
    ybar,  
    title style = {text width = 9cm,align = center},
    title = {\footnotesize Q9: Do you prefer having a simulator capable of modelling every type of blockchain?},
    ylabel={\footnotesize Frequency}, 
    xlabel={\footnotesize Agreeing level},  
    symbolic x coords={Strongly disagree, Disagree, Neutral, Agree, Strongly agree}, 
    bar width=1cm,
    xtick=data,  
     nodes near coords, 
    nodes near coords align={vertical},  
enlargelimits=true,x tick label style={font=\footnotesize,text width=1cm,align=center},
    y label style={below=0.5mm},
    x label style={below=2mm},
    bar width=0.8cm,                
    height=5.7cm,
          width=8.5cm,
          style={xshift=0pt,yshift=0pt,anchor=north,font=\footnotesize}
    ]  
\addplot +[black, pattern=north east lines] coordinates {(Strongly disagree,3) (Disagree,7) (Neutral,9) (Agree,4) (Strongly agree,2) };  
  
\end{axis}  
\end{tikzpicture}  
    \caption{Participants' thoughts about modelling different blockchain types in the simulator.}
	\label{types}
\end{figure}
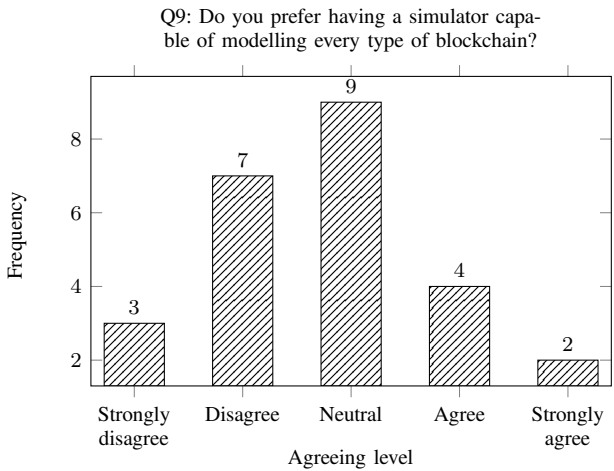

Finally, given that there are numerous types of blockchain, there is a need to comprehend if it is essential to have a simulator that can model the diverse types. Accordingly, the participants were asked about this, with their responses to this question are presented in Figure~\ref{types}. According to the participants’ perspectives, the majority (nine participants, 36\%) are neutral towards this. Alternatively, four participants (16\%) agreed, while two participants (8\%) strongly agreed. Finally, ten participants (40\%) either strongly disagreed or disagreed with this notion.

For the picture to be complete, Table \ref{mtachobj} matches the questionnaire questions to the predefined objectives.

\begin{table}[]
    \centering
    \caption{Matching the questionnaire questions to the predefined objectives. For the question numbering, please refer to Figures \ref{IoTfamiliarity} - \ref{types}.}
    \scalebox{0.8}{

    \begin{tabular}{c||cccccc||cc||c}
\toprule 
 Question  & \multicolumn{6}{c||}{Objective 1} & \multicolumn{2}{c||}{Objective 2} & Objective 3 \\\cline{2-9}
 \#& a & b & c & d & e & f&a & b &  \\
\midrule 
 Q1 & \cmark &  &  &  &  &&  &  &  \\
Q2 &  & \cmark  &  &  &  &&  &  &  \\
Q3 & \cmark & \cmark  &  & & &  &  &  &  \\
Q4 &  &   &  &  &  &  & \cmark && \cmark \\
Q5 &  &   & \cmark &  &  &&  &  & \cmark \\
Q6 &  &   &  & \cmark &  &&  &  & \cmark \\
Q7 &  &   &  &  & \cmark &&  &  & \cmark \\
Q8 &  &   &  &  &  &  \cmark&  && \cmark \\
Q9 &  &   &  &  &  &  &&  \cmark& \cmark \\
 \bottomrule
\end{tabular}
}
    \label{mtachobj}
\end{table}
\subsection{Interviews}
\label{InterviewsResult}
Interviews were conducted with a set of participants to collect information concerning their opinions on using IoT-based Blockchain, as well as comprehending their requirements for the simulation software for assessing blockchain-based IoT. The participants’ responses were assessed from active, analytical, and critical perspectives, with their suggestions being clarified. Three questions were posed:

\begin{itemize}
    \item \textbf{What are the major challenges you face when dealing with blockchain-based IoT for any evaluation purposes?}
    \item \textbf{Which features make blockchain suitable for the IoT?}
    \item \textbf{What are the anticipated outcomes of utilising blockchain within the IoT?}
\end{itemize}

\textbf{P1} stated that \textit{``There are many challenges based on the current proposed model. The obstacle lies in investigating the performance and cost of these technologies. Also, there are many proposed simulators for Blockchain and IoT in the literature; however, each simulator either focuses on IoT or blockchain. As a researcher, I prefer having a multi-discipline simulator that can simulate IoT devices in sensing and sending data to the edge/fog layer then to cloud, while using blockchain in different layers"}. Regarding the second question, he remarked, \textit{``The majority of IoT applications such as healthcare data is of high importance and needs to be securely handled. I believe blockchain is a strong fit for this scenario because of its features (for example, decentralisation) that dispense a third party to manage data"}. Lastly, for the third question, he suggested that \textit{``With the rapid development of IoT technology and the large number of devices expected to be connected, I believe blockchain would alleviate security issues. For example, identity management and access to the IoT should be more secure and trusted, using a reliable tool for controlling data access"}.

\textbf{P2} stated concerning the first question, "The main challenge I faced with the IoT and blockchain technologies is the difficulty of monitoring systems’ performance. \textit{``The challenge is that it does not cover all of my required features. I often use a cloud simulator to evaluate the system. Having a Blockchain simulator with IoT features that can track every transaction and system throughput will ease my tasks. This could become an efficient simulator, utilising both blockchain and IoT power"}. Concerning the second question, the participant explained that \textit{``not all the IoT data are of high importance, but there is still a need to secure the sensitive data and enhance privacy"}. Finally, he stated that \textit{``I believe blockchain can mitigate several of the IoT issues related to privacy. Also, blockchain can define a set of policies needed to control IoT data access"}.

In reply to the first question, \textbf{P3} mentioned that \textit{``One of the most important blockchain-based IoT challenges is system evaluation because of the heterogeneity and mobility of IoT devices. Personally, I prefer to assess the system from different viewpoints, ranging from general performance (computational time, transaction latency and throughput) to security, but there is no simulator permitting this"}. Responding to the second question, he said that \textit{``Data storage is a crucial metric to determine the applicability of blockchain with the IoT. The IoT devices sense the environment and send data in real-time. This implies that we have plenty of data per second. Accordingly, blockchain cannot be used as data storage for all data. Hence, I prefer storing only the most important data; I think that this can be reliable"}. Concerning the third question, he asserted that \textit{``Every single device can be identified using a permissioned blockchain network that is used by all parties involved. This implies that data is generated by an identified device (trusted), in the sense that the generated data has a unique identification number, hence ensuring immutability. In this scenario, it could be appropriate to track the data in the supply chain context"}.

\textbf{P4} noted concerning the first question that \textit{``The challenge is how to obtain various statistics about the system, like the number of generated transactions, number of blocks and time of confirmation, both for the block and transaction. These metrics give me an indicator about the proposed system, which is essentially the same as the real world and enables me to make decisions"}. Regarding question two, the participant stated that \textit{``IoT data can be immutable and distributed over time in the blockchain network. Participants in the blockchain network can ensure the data's authenticity and that it will never be tampered with"}. Finally, concerning question three, he remarked that \textit{``I advise using blockchain to keep sensitive IoT data where security is ensured. Also, as IoT devices are the data source, there is a need for reliable analysis which will not be carried out if there are no device management criteria. I believe this can be carried out by blockchain, for example, using smart contracts"}. 

\textbf{P5} commented that \textit{``Assessing the system performance is very important. In the context of blockchain and the IoT, it is difficult to measure performance without simulation due to the complexity of both technologies. So, from my point of view, the simulator enables me to test the system from diverse aspects. Specifically, I can configure the number of IoT devices and protocols used, while at the same time determine the size of transactions, either for blockchain or the IoT (end to end)"}. Regarding the second question, he suggested that \textit{``One of the advantages of blockchain is decentralisation, as it can prevent a single point of failure and bottlenecks from occurring. I see that blockchain benefits the IoT by ensuring reliable data transfer"}. Finally, he stated \textit{``I believe that blockchain would provide the IoT developers with more secure solutions due to its features"}.

Concerning the first question, \textbf{P6} remarked that \textit{``The challenge lies in determining if the simulator supports more than one measure, such as latency, throughput, total time, along with the number of blocks created to analyse the overall performance of blockchain and the Internet of Things. Based on my experience, it is difficult to cover all aspects of the IoT and blockchain simultaneously, but the simulator can cover the general aspects of both technologies in different scenarios"}. Concerning the second question, he stated that by and large \textit{``Due to the limited processing capabilities of IoT devices, third-party service providers are generally used to process additional data. By entrusting sensitive user data to third-party service providers, users must trust data protection and privacy. This trust coincides with the danger of breaking data privacy and policies. Blockchain’s traceability can help in these situations"}. Finally, he expressed that \textit{``Blockchain is a promising choice when it comes to ensuring privacy and applying security"}. 

\section{Discussion}
\label{discussion}
\subsection{Questionnaire}
\label{questionnaireDiscussion}
As previously stated, we used a questionnaire to gather the perspectives of 25 individuals who are very acquainted with IoT and blockchain, as shown by their responses to questions ~\ref{IoTfamiliarity} and ~\ref{Blockfamiliarity} , respectively. The participants believe in blockchain as a means of mitigating the problems facing the IoT, as depicted in question~\ref{thoughts}. This aligns with the results of the first two questions. As previously said, IoT applications will become more important for individuals and organisations in the next years, despite the fact that they are now struggling with concerns such as trust, reliability, security, and performance. Blockchain technology incorporates a range of features particularly with respect to security, and IoT application builders, hope that Blockchain may mitigate the aforementioned concerns. The participants expressed a positive attitude toward having an integrated simulator to mimic the behaviour of IoT and blockchain technologies, which is illustrated in question~\ref{configuation}. Specifically, there is a requirement for a simulation tool that can be adopted to  assist with modelling applications that integrate Blockchain with IoT. The majority of participants were against storing all of the sensed/gathered data in blockchain, as their responses to question~\ref{IoTdata} showed. The reasons underpinning this attitude are that firstly, the IoT senses and collects a tremendous amount of data per second, while secondly, blockchain is acknowledged as suffering from scalability issues. Accordingly, specialists may reject this idea to avoid blockchain failure. Conversely, the majority agree with the benefits of having multiple consensus algorithms, as depicted in question~\ref{consensus}. This is because multiple consensus algorithms have been developed in the literature, each having its own procedures and applications. Consequently, as the applications change, the required consensus algorithm may change; this is why different consensus algorithms are necessary in the simulator. Another notable point is that the majority of participants prefer capabilities for in-depth investigation of the blockchain logs as reflected in question~\ref{log}. Specifically there is a need for carrying out deep tracking of transactions to investigate potential delays and other performance issues. Additionally, there is a high degree of agreement on using IoT devices as blockchain nodes, as established by question~\ref{pi}. This is intuitive because IoT micro-elements have reasonable processing power while being small in size. Finally, in question~\ref{types}, participants were not in favour of having a simulator comprising various blockchain types. The reason behind this was that each type has its own features. Hence, implementing all of these features in a single simulator may increase complexity which, in turn, limits its applicability. By and large, from the questionnaire, we may deduce the necessity of developing a simulator that can mimic the integration of blockchain with the IoT.

\subsection{Interview}
\label{interviewDiscussion}
For a more in-depth understanding, online interviews with specific participants were organised. The interview comprised three questions. As mentioned previously in section~\ref{InterviewsResult}, the first question focused on understanding the significant challenges facing the participants. We analysed the six participants’ responses and determined that performance assessment is the principal challenge. This is because they must ensure their developed systems’ smooth performance before initiating their actual implementation.

Furthermore, there is a clear focus on the need to analyse performance alongside ensuring safety and data privacy. However, the existing simulators have their own limitations. The main one is that only one or two aspects are implemented in each simulator. Almost all participants lacked the simulator software to assist with this task. Regardless of a large number of simulators being developed in the existing literature, no extensive simulator has been developed that can focus on analysing the system from various standpoints. While agreeing to align blockchain with the IoT, the fundamental problem that hinders the participants is how to assess IoT systems with blockchain features. Unfortunately, this restricts blockchain’s applicability with IoT. Moreover, the participants observed that most of the existing simulators do not support a wide range of parameter settings, thus limiting the evaluation to a certain viewpoint while neglecting other informative perspectives. Additionally, this reflects their intuition toward comparing the developed system with similar systems to obtain an improved understanding of the system. In this regard, they all expressed optimism about having a simulator to facilitate their task. 

Through analysing the participants’ responses, we observed that data security and privacy are the most important features; each of the participants concurred that the majority of the IoT collected data could be exceedingly sensitive and must be handled carefully. Further, the participants’ responses conveyed how security is a worrying issue related to the IoT. A further point is that dependency on a third party to ensure security is usually accomplished using servers. Conversely, the concept which has made blockchain attractive for dealing with data, namely that it is tamper proof, has gained considerable attention from the participants. Accordingly, they anticipate using blockchain to store IoT data to secure it. This will ensure that the sensed data will never be reverse tampered. Even so, this was a controversial point, with certain participants mentioning that blockchain has shortcomings in terms of scalability. Specifically, massive amounts of data are gathered from IoT devices per second. Therefore, storing this extraordinary volume of data in blockchain may fail. Thus, a trend has emerged of keeping only the most valuable data in the chain. Notably, an additional prominent blockchain feature concerning the IoT is the decentralisation concept. 

Finally, responding to the third question, it was apparent that the participants had considerable confidence in blockchain as an efficient alternative in relation to the IoT. Without exception, all of the participants believed that blockchain offers a lot of potential to address many of IoT's difficulties. Specifically, they mentioned that using blockchain could easily ensure IoT’s data security and privacy. For example, IoT has no clear device management policies. It is not a straightforward task to define who can access one’s device and when. But this is no longer an issue when using blockchain, because devices can be registered as blockchain nodes. Therefore, they will have a clear management role. A further point is that for very complex applications it can be very difficult to track IoT data along its journey. This is due to the potentially large number of devices involved. Conversely, it will become far simpler to track IoT data using blockchain, where data and transactions are always time-stamped and up to date, therefore users can query a product’s status and location at any point in time.

\section{Recommendations}
\label{recommendation}
The results presented in the previous sections have evidenced a broad belief that blockchain can benefit IoT applications and enhance its applicability by alleviating its limitations. Moreover, the majority of participants in our studies agreed that it is necessary to have an integrated blockchain IoT simulator to aid in the development and evaluation of application that integrate blockchain and IoT technologies. On this basis, we recommend greater research and exploration of the design and development of an integrated blockchain IoT simulator. Considering the lack of such a simulator in the literature, this calls for greater research and the need to attract the attention of contemporary researchers. 

\section{Conclusions}
\label{conclusion}
IoT systems are becoming increasingly widespread. Yet, because they are centralised, they have their own set of constraints. However, it is anticipated that blockchain has the potential to unlock new opportunities for IoT. A major problem is that there is no credible simulator for evaluating blockchain as a solution for IoT problems. This motivates our current work to shed light on the design of such a simulator. To better understand this notion, two studies were conducted involving a questionnaire and interviews, respectively, involving a number of experts. The results of the questionnaire indicate considerable familiarity with IoT and blockchain. The results also demonstrated a strong belief in blockchain as a technology that may mitigate a number of IoT’s issues. This was further confirmed by the results of the interviews with the experts. Moreover, from the questionnaire and interview results, we discovered that a primary challenge is the lack of simulator software with the ability to replicate the behaviour of IoT applications combined with blockchain. In this context, the aim of our future work is to provide software that can simulate IoT integration with blockchain, allowing systems to be evaluated and validated before they are deployed in the real world.

\bibliographystyle{IEEEtran}
\bibliography{IEEEabrv,References}
\end{document}


\maketitle
\section{Questionnaire}
\begin{figure}[ht]
    \centering
    
    \includegraphics{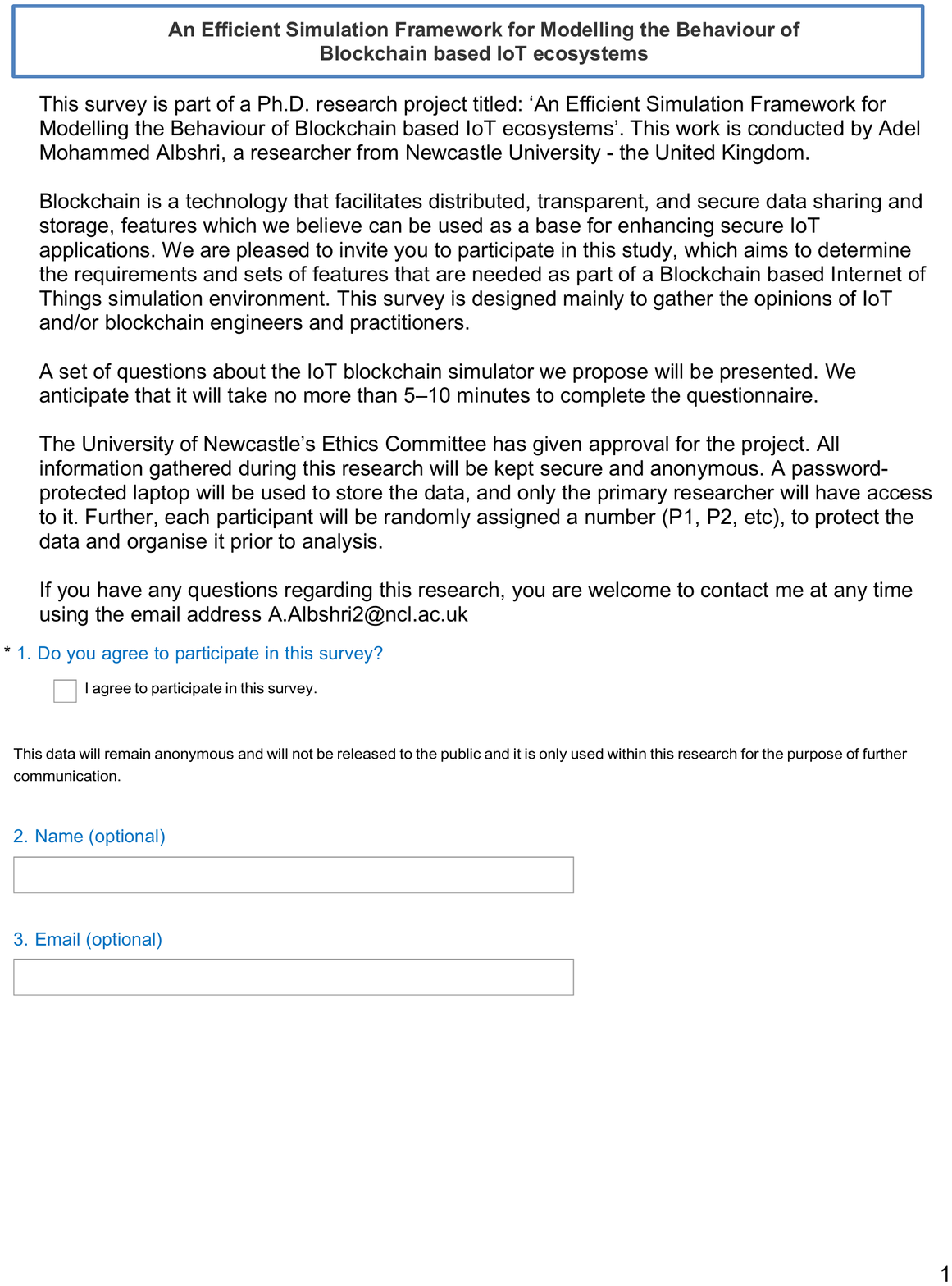}
    \caption{Caption}
    \label{fig:my_label}
\end{figure}
\section{Statistical Results of Questionnaire}
\section{Interview}